%% file: mmv_ra.tex
\documentclass[conference]{IEEEtran}   
\usepackage[english]{babel}
\usepackage{amssymb}
\usepackage{amsmath}
\usepackage{amsthm}
\usepackage{epsfig}
\usepackage{graphicx}
\usepackage{color}
\usepackage{dsfont}
\usepackage{cleveref}
\usepackage{subcaption}
\usepackage{enumerate}
\usepackage{algorithm}
\usepackage{algpseudocode}
\usepackage[backend=bibtex,sorting=none]{biblatex}
\newenvironment{Figure}
  {\par\medskip\noindent\minipage{\linewidth}}
  {\endminipage\par\medskip}
    
\usepackage{tikz}
\usetikzlibrary{positioning}
\usetikzlibrary{arrows,chains, shapes, shapes.misc}
\usepackage{pgfplots}
\pgfplotsset{%
    ,compat=1.12
    ,every axis x label/.style={at={(current axis.right of origin)},anchor=north west}
    ,every axis y label/.style={at={(current axis.above origin)},anchor=north east}
    }

\input{macros}

\DeclareMathOperator*{\argmin}{arg\,min}

\begin{document}

\newcommand\figref{Figure~\ref}

\newcommand{\ben}{\begin{enumerate}}
\newcommand{\een}{\end{enumerate}}

\newcommand{\beq}{\begin{equation}}
\newcommand{\eeq}{\end{equation}}

\newcommand{\bi}{\begin{itemize}}
\newcommand{\ei}{\end{itemize}}

\newcommand{\1}{\mathds{1}}
\markright{\today}

\IEEEoverridecommandlockouts

\title{Massive MIMO Unsourced Random Access}
\author{\IEEEauthorblockN{Alexander Fengler, Giuseppe Caire, Peter Jung, and Saeid Haghighatshoar}
\IEEEauthorblockA{Communication and Information Theory Chair, Technische Universit\"at Berlin}
}

\maketitle

\begin{abstract}
We consider an extension of the  massive unsourced random access originally proposed in 
\cite{Polyanksiy:isit17} to the case where the receiver has a very large number of antennas  (a massive MIMO base station)
and no channel state information is given to the receiver (fully non-coherent detection). 
Our coding approach borrows the concatenated coding idea from  \cite{Ama2018}, combined with 
a novel non-Bayesian ``activity detection'' algorithm for massive MIMO random access channels, that outperforms currently proposed
Bayesian vector AMP (VAMP) schemes currently proposed for activity detection, and does not suffer from 
the numerical instabilities and requirement for accurate a priori statistics as VAMP. 
We show that the required transmit $E_b/N_0$ for reliable communication 
can be made arbitrarily small as the number of receiver antennas $M$ grows sufficiently large. 
\end{abstract}


\section{Introduction}

In some Internet of Things applications, it is envisaged that a very large number of objects send sporadic data 
via a wireless channel to a central collector. Such objects may be sensing-enabled appliances, 
that are mass-produced and disseminated into the environment without any 
specific centralized control. 
For massive scalability and production costs reasons, the transmission protocol (including signaling and channel coding)
shall be hard-wired and identical to all devices. This poses a new random access problem 
known as ``unsourced massive random access'', where the goal of the 
receiver (data collector) consists of decoding the messages transmitted by a small
number of active nodes on each transmission slot, whose identity is not known a priori, and 
such that these nodes make use of exactly the same channel codebook (or, more in general, the same transmission protocol).
In this context, if senders want to identify themselves, they can include their ID into the 
information message itself. Therefore, the goal of the receiver is to decode the {\em list} of active user messages up to permutations.

This new information theoretic problem has  been posed by Polyanskiy in  \cite{Polyanksiy:isit17}. In Polyanskiy's formulation, 
the number of users is comparable to the channel block length (e.g., imagine a city-wide IoT network with
$\sim 10^6$ sensors, each of which sends sporadic data using a codebook of block length $\sim 10^4$). 
In this regime, under the classical notion of probability of error used in the information theoretic multiple access channel \cite{cover2006elements}, 
no reliable communication is possible. Hence, Polyanskiy proposes 
the Per-User Probability of Error  (PUPE), i.e., the average fraction of mis-decoded messages over the number of active users, 
as a more practically meaningful performance metric. In  \cite{Polyanksiy:isit17} the channel is modeled as a real adder with additive white Gaussian 
noise (AWGN), where the discrete-time baseband signal received at the decoder is given by 
\begin{equation} \label{real-adder}
y_t = \sum_{k \in \Kc_a} x_t(m_k) + z_t,  \;\;\;\; t = 1, \ldots, n,
\end{equation}
where $\Kc_a$ is the set of $|\Kc_a| = K_a$ active users in a population $\Kc_{\rm tot}$ of $|\Kc_{\rm tot}| = K_{\rm tot}$ total users, 
$x_t(m)$ is the $t$-th symbol of the $m$-th codeword of a common codebook $\Cc \subset \RR^n$ of cardinality $2^{n R}$, 
$m_k \in [1 : 2^{nR}]$ is the message transmitted by active user $k \in \Kc_a$, 
and $z_i \sim \Nc(0, N_0/2)$ is the real AWGN.  For this channel,  \cite{Polyanksiy:isit17} established quite tight achievability and converse bounds
to the minimum energy per bit over $N_0$ ($E_b/N_0$) necessary for reliable communication.  

The sporadic communication patterns and the very large number of potential users $K_{\rm tot}$ rule out most of the
current network solutions, which are essentially ``grant-based'' random access. 
In such schemes, when a user wishes to send a data packet, it must first access a dedicated Random Access Control Channel (RACCH), 
identify itself, and ask for a transmission resource. After being granted access, the user sends its message on 
the allocated dedicated resource (e.g., a combination of time-frequency slot and space beam, in modern MIMO-enabled space-division multiple access). 
During the granting process, each user makes use of some unique feature in its signaling scheme (e.g., a unique combination of 
pilot/signature signal and time-frequency slot) such that the base station is aware of the identity of the user requesting access. 
Furthermore, during this hand-shaking procedure, pilot signals are used to estimate the channel state (e.g., channel vectors
in fading MIMO communications), such that channel-state based precoding/beamforming is made possible. 

In contrast, it is becoming more and more apparent that 
radically new grant-free strategies must be developed for 
for the unsourced massive random access scenario described before, since in this scenario the protocol overhead cost
(in rate and energy) for grant-based communication and the inconvenience of individually distinct codebooks at the user devices would be
simply too large. Hence, in this scenario the active users have to be simultaneously detected and decoded within a single communication
phase.

Providing {\em explicit} coding schemes to approach the random coding achievability results 
is an interesting and relevant challenge for the design of practical IoT systems. In fact, for the 
unsourced massive random access setting described above, the gap in $E_b/N_0$ between random coding achievability
and the state of the art random access schemes known to date is very large. Therefore, a large gain 
in energy efficiency can be achieved through an improved coding design. 

Motivates by Polyanskiy's results,  \cite{Ama2018} proposed a coding scheme for the real-adder AWGN channel
(\ref{real-adder}), based on partitioning the transmission slot into subslots, and letting each active user send a codeword from 
a common codebook across the subslots. The common codebook is obtained by concatenating 
an outer tree code with an inner ``compressed sensing code''. The inner encoder maps each submessage into one column of a
given (real) coding matrix. The inner decoder must identify which columns of the matrix have been transmitted from the received 
noisy  superposition. This is a classical ``sparse support identification'' problem, well investigated in the compressed sensing (CS) literature. 
In particular,  due to the fact that the support is a 0-1 sparse signal,   \cite{Ama2018} makes use of the 
Non-Negative Least-Squares (NNLS) approach, which is known to be computationally very efficient and well-suited for
the sparse support identification problem \cite{kueng2018robust,slawski2013non}. 
The inner decoder produces a sequence of active submessages lists across the
subblocks. The task of the outer tree decoder is to  ``stitch together'' the
submessages such that each sequence of submessages is a valid path in the tree.

For the simple real-adder AWGN channel model (\ref{real-adder}), the aspect of channel estimation is absent
since the channel coefficients are all fixed to 1. Hence, detection is (implicitly) coherent by default. 
In practice, through, wireless communication is affected by channel pathloss attenuation, phase shifts, 
and fading. In particular, when extending the model (\ref{real-adder}) to the case of a multiple-antenna base station receiver, 
it is completely unreasonable to assume that the channel matrix between the $K_a$to active users and the 
$M$ receiver antennas is known a priory, and often even that the channel pathloss coefficients (i.e., the active user signal strengths) are known
a priori. In standard multiuser MIMO communications it is usually assumed that the uplink data transmission comes with
some dedicated pilot field, where each user is given an individual pilot sequence, such that the base station receiver can 
estimate the corresponding column of the channel matrix and use a ``coherent'' approach for detection. 
Pilot-based coherent detection is widely studied and ubiquitously used in the massive MIMO literature 
\cite{marzetta_larsson_2016}.
It is obvious that, in the unsourced massive random access context, assigning a dedicated pilot to each user violates the 
``single common codebook'' paradigm. 

In parallel to the line of work on unsourced massive random access, another line of work has focused on 
activity detection with massive MIMO reception \cite{liu2018sparse,liu2018massive,chen2018sparse,wang2018fly,Hag2018}.
In this context, users are given individual pilot sequences and the superposition of the active users' pilot sequences is observed at the output ($M$ antenna ports) 
of a massive MIMO receiver. The goal of the receiver consists of identifying the set of active users and, in certain applications, also estimate their large-scale
channel pathloss coefficients (i.e., their signal strength).  
Activity detection is similar to widely studied problems in many other fields such as network tomography and neuron activation in 
brain imaging. In the context of wireless communications it has been intensively investigated for grant-free random access
\cite{liu2018sparse} and it is obviously a prototypical application of CS.  
In \cite{liu2018sparse,liu2018massive} it is assumed that a slot is divided into two subslots, one for channel estimation, and the other for
data detection. Activity detection takes place on the channel estimation subslot, where users send individual (unique) pilot sequences. 
It follows that the above mentioned works on activity detection do not apply immediately to the unsourced massive random access paradigm. 

An interesting approach that moves activity detection in the direction of unsourced massive random access is taken in \cite{larsson2012piggybacking,senel2017device}, where the same activity detection approach of \cite{liu2018massive,chen2018sparse} is 
used to send information in a piggybacking mode: if each user is given a set of possible pilot sequences instead of just a single one, 
using the CS-based activity detection scheme it is possible not only to identify the active users, but also which codeword (pilot sequence) each active user 
has sent. Thus, each active user can send information together with its activity. In addition, the CS scheme of  \cite{liu2018massive,chen2018sparse} is used 
in \cite{larsson2012piggybacking,senel2017device} as a non-coherent decoder, i.e., there is no need of explicit MIMO channel matrix estimation before 
detection. Nevertheless, the piggybacking scheme of \cite{larsson2012piggybacking,senel2017device} is still based on the fact that each user has a unique 
individual codebook, and therefore it is not compliant with the unsourced massive random access paradigm. Furthermore, in order to send a 
$B$ bit message each user needs a codebook of $2^B$ pilot sequences. Therefore, the scheme of  \cite{larsson2012piggybacking,senel2017device} 
can only work for messages of very small size (e.g., for $B$ as small as 100 bits, the scheme would be completely impractical). 

\subsection{Our Contribution}

In this work we consider the unsourced massive random access problem for the case where the 
base station has a massive number of antennas. We refer to this scenario as the 
{\em Massive MIMO Unsourced Random Access} channel. We combine the concatenated approach of \cite{Ama2018}
with a novel non-coherent massive MIMO activity detection scheme first proposed by the authors in \cite{Hag2018}. 
The novel activity detection algorithm is  non-Bayesian, and treats the activity detection pattern including 
the users' large-scale pathloss coefficients as a deterministic unknown sparse non-negative vector.
This is different from the Bayesian Vector Approximated Message Passing (VAMP) approach taken in  
\cite{liu2018massive,chen2018sparse,larsson2012piggybacking,senel2017device}, where the pathloss coefficients are considered either 
fully known or statistically fully known. Both such assumptions are highly unrealistic in 
the context of massive random access. The knowledge of the channel pathlosses for all $K_{\rm tot}$ users implies that at the time of deployment
(imagine millions of sensors!) each channel is calibrated and the attenuation due to the propagation pathloss is stored in memory. 
Furthermore, such calibration procedure must be repeated over time since the propagation conditions change. 
Now, for large-scale effects such as shadowing, statistical fluctuations of the channel pathloss of up to 6-8 dB are possible, with
over a time scale (coherence time) from tens of seconds to tens of minutes (imagine traffic sensors in the presence of moving large objects such as 
trucks and buses, motion of tree foliage and other large-scale effects of this sort). It is easy to figure out that maintaining calibration 
for a very large network would be very hard and impractical. Furthermore, even the statistical knowledge of such coefficients is difficult to 
acquire in practice. In fact, distance dependent pathloss laws are somehow hard simplifications of reality, and are maybe useful for 
validating performances via simulation, but definitely not accurate and reliable enough to be exploited as statistical information in a Bayesian detection context.
In practice, extracting such statistical information is tantamount collecting a large sample of the receive signal energy from each node in the network, 
and forming a combined histogram for all $K_{\rm tot}$ users. Again, this requires a lot of data exchange which eventually is similar to the 
calibration said before.  

Perhaps even most surprisingly, our proposed non-Bayesian activity detection algorithm outperforms the Bayesian 
VAMP even under the very favorable assumption (for VAMP) of fully known pathloss coefficients (let alone the case of statistically known). 
Furthermore, VAMP shows a very critical ``unstable'' behavior when the number of base station antennas is very large
with respect to the pilot dimension. In contrast, our algorithm uniformly improves as the number of antennas grows large
for any fixed pilot dimension, and shows no numerical instabilities. Also from a computational complexity viewpoint, our scheme is 
only slightly more complex than VAMP. Therefore, it is definitely the algorithm of choice for this application, requiring much less a priori knowledge, 
and offering significantly better performance in the relevant regime of large number of antennas, and similar complexity. 
 
To our knowledge, the scheme proposed in this work is the first practical coding solution to the unsourced 
massive random access problem as posed in \cite{Polyanksiy:isit17}
for a massive MIMO block-fading channel with noncoherent decoding. 
In our scheme, the inner code essentially reduces the MIMO random access to a random ``OR-channel'' to the outer code, where 
each signal dimension at the output of the outer code is associated with ``activity'' or ``non-activity'', whether one or more messages
are active or no message is active on that dimension.  This is repeated for each subslot, such that the sequence of 
lists of active messages over the subslots is passed to the outer tree code as in  \cite{Ama2018}.

Our numerical simulations suggest that the proposed scheme performs substantially better then previous approaches. 
Furthermore, we show that theoretically that the required transmit $E_b/N_0$ for reliable communication 
can be made arbitrarily small as the number of receiver antennas $M$ grows sufficiently large. 

The structure of the papers is follows: we will introduce in Section
\ref{sec:sysmodel} the system model on more detail. Then, in Section
\ref{sec:codingscheme} a inner and outer encoding is proposed and
efficient decoding algorithm is presented. We will discuss our setup
further in Section \ref{sec:analysis} and show first numerical
experiments in Section \ref{sec:numerics}.

\section{System Model}
\label{sec:sysmodel}

We consider the block fading Gaussian MAC with $M$ receive antennas and $K_a$ out of $K_{\rm tot}$ active
users, where each user is employing the same
codebook $\mathcal{C} \subset \CC^n$. The received $t$-th signal sample at the $i$-th antenna is given by: 
\beq
y_{t,i} = \sum_{k \in \mathcal{K}_a} \sqrt{g_k} h_{k,i} x_t(m_k) + z_{t,i},
\label{eq:channel}
\eeq
for  $t = 1, \ldots, n$ and $i = 1, \ldots, M$. As before, $x_t(m)$ denotes the $t$-th symbol of the $m$-th codeword of a common  
codebook $\Cc = \{\xv(m) : m \in  [1 : 2^{nR}] \}$, $m_k$ is the message transmitted by active user $k \in \Kc_a$,
$g_k \in \RR^+$ is the large-scale channel gain coefficient of user $k$ and $h_{k,i} \in \CC$ is the small-scale channel coefficient 
from user $k$'s transmitter to the base station antenna $i$. In this complex baseband model the noise $z_{t,i}$ is i.i.d. $\sim \Cc\Nc(0,N_0)$
(circular symmetric Gaussian), independent across time and antennas. 
The codebook $\Cc$ satisfies the per-message power constraint $(1/n) \sum_{t=1}^n |x_t(m)|^2 \leq P$ for all $m$.

We also assume that the MIMO channel is spatially white, zero-mean Rayleigh fading, such that the coefficients $h_{k,i}$'s are also 
i.i.d. $\sim \mathcal{CN}(0,1)$. Furthermore, we assume a block-fading model for which the coefficients 
$h_{k,i}$ are random but fixed for all coding dimensions $t = 1, \ldots, n$. Notice that such dimensions can be obtained in the time-frequency domain. 
Therefore, the model (\ref{eq:channel}) applies to both narrow-band frequency-flat fading and to wideband frequency-selective 
fading with OFDM modulation. Furthermore, the assumption of constant small-scale fading over the whole transmission duration can be relaxed
to constant over much smaller subblocks, as explained in the following. 

Given that each user uses the same codebook, it is inherently impossible for the receiver to
identify users and assign messages to users. Instead, the receiver produces a list of
decoded messages $\mathcal{L}$ and the performance is expressed in terms of the 
{\em Per-User Probability of Misdetection}, defined as the average fraction of 
transmitted messages not contained in the list, i.e., 
\beq
p_{md} = \frac{1}{K_a}\sum_{k \in \mathcal{K}_a} \PP(m_k \notin \mathcal{L}),
\eeq
and the {\em Per-User Probability of False-Alarm}, defined as the average fraction of decoded messages that 
were indeed not sent, i.e., 
\beq
p_{fa} = \frac{|\mathcal{L}\setminus \{ m_k : k \in \Kc_a \}|}{|\mathcal{L}|}.
\eeq
Notice that the error probabilities are independent of the total number of users $K_{\rm tot}$ and
depend only on the cardinality of the (random) active set.  
The power efficiency of the decoding scheme is measured in terms of
$E_b/N_0:= \frac{P}{R N_0}$.

\section{Proposed Coding Scheme}
\label{sec:codingscheme}

As anticipated before, we follow the concatenated coding scheme approach of  \cite{Ama2018}, suitably adapted to our case. 
Let $b = nR$ denote the number of bits per user message. For some suitable integers $L \geq 1$ and $J > 0$, we
divide the $b$-bit message into blocks of size $b_1, b_2, \ldots, b_L$ such that $\sum_\ell b_\ell = b$ and such that
$b_1 =  J$ and $b_\ell < J$ for all $\ell = 2, \ldots, L$. 
Each subblock $\ell = 2, 3, \ldots, L$ is augmented to size $J$ by appending $p_\ell = J - b_\ell$ parity bits,  
obtained using pseudo-random linear combinations of the information bits of the previous blocks $\ell' < \ell$. 
Therefore, the sequence of coded blocks forms a tree of depth $L$. 
The pseudo-random parity-check equations are identical for all users, i.e., each user makes use exactly of the same outer code.
For more details on the outer coding scheme, please refer to  \cite{Ama2018}. 

\subsection{Inner Code}

In this section we focus on the inner code. Given $J$ and the subslot length $n_0 = n/L$, the inner code is used to 
transmit in sequence the $L$ (outer-encoded) blocks. 
Let $\Am \in \CC^{n_0 \times 2^J} = [\av_1,...,\av_{2^J}]$,
be a matrix with columns normalized such that
$\|\av_i\|^2\leq n_0 P$. Each column of $\Am$ represents one {\em inner} codeword.
Each active user $k \in \Kc_a$, let $i_k(1), \ldots, i_k(L)$ denote the sequence of $L$ (outer-)encoded 
$J$-bit messages produced by the outer encoder, represented as integers in $[1:2^J]$. 
Then, each user $k \in \Kc_a$ simply sends in sequence, over consecutive subblocks of length $n_0$, 
the columns $\av_{i_k(1)}, \av_{i_k(2)}, \ldots, \av_{i_k(L)}$ of the coding matrix $\Am$. 
From (\ref{eq:channel}), collecting the $n_0$ channel uses forming subslot $\ell$ and the outputs of all $M$ antennas 
into a $n_0 \times M$ matrix $\Ym_\ell$, the received signal is given by 
\begin{equation} 
\Ym_\ell = \Am \Bm_\ell \Gm^{1/2} \Hm + \Zm_\ell, \;\;\; \ell = 1, \ldots, L,
\label{matrix-channel}
\end{equation}
where $\Gm = \diag(g_1, \ldots, g_{K_{\rm tot}})$, 
$\Hm \in \CC^{K_{\rm tot} \times M}$ with Gaussian i.i.d. entries $\sim \Cc\Nc(0,1)$, 
$\Zm_\ell \in \CC^{n_0 \times M}$ has Gaussian i.i.d. entries $\sim \Cc\Nc(0,N_0)$,
and $\Bm_\ell \in \{0,1\}^{2^J \times K_{\rm tot}}$ is a binary activity matrix with columns 
$\bv_k^{(\ell)}$ such that: i) if $k \notin \Kc_a$, then $\bv_k = \zerov$, and ii) if 
$k \in \Kc_a$, then $\bv^{(\ell)}_k = \ev_{i_k(\ell)}$, the standard basis vector in dimension $2^J$, 
with all-zeros and a single 1 in position $i_k(\ell)$. Notice also that, for what follows, 
we may equivalently assume that $\Hm$ changes at each subslot $\ell$, i.e., replace $
\Hm$ in (\ref{matrix-channel}) with $\Hm_\ell$ was dependent on $\ell$. The results will remain the same
provided that the marginal statistics of the matrices $\Hm_\ell$ is Gaussian i.i.d. as said above (independently of their correlation across the 
subslots). 


Let's focus on the matrix $\Xm_\ell =  \Bm_\ell \Gm^{1/2} \Hm$ of dimension $2^J \times M$. 
The $r$-th row of such matrix,\footnote{We use underline to denote row vectors, while simple bold-face small-case letters denote standard column vectors.} 
denoted by $\underline{\xv}^{(\ell)}_r$, is given by 
\beq
\underline{\xv}^{(\ell)}_r = \sum_{k\in\mathcal{K}_a}\sqrt{g_k} b^{(\ell)}_{r,k} \underline{\hv}_k, 
\eeq
where $\underline{\hv}_k$ is the $k$-th row of $\Hm$, and  where $b^{(\ell)}_{r,k}$ is the $(r,k)$-th element of 
$\Bm_\ell$, equal to 1 if the active user $k$ sends subblock number $r$ (i.e., if $i_k(\ell) = r$) and zero otherwise.
It follows that $\Xm$ is Gaussian with independent entries $\sim \Cc\Nc\left (0, \sum_{k \in \Kc_a} g_k b^{(\ell)}_{r,k} \right )$. 
Since the submessages are independently and uniformly distributed over $[1:2^J]$, 
the probability that the $r$-th row is identically zero is given by $(1 - 1/2^J)^{K_a}$. 
Hence, for $2^J$ significantly larger than $K_a$, the matrix $\Xm$ is row-sparse. 


Let $\gamma^{(\ell)}_r := \sum_{k\in\mathcal{K}_a}g_k b^{(\ell)}_{r,k}$ and
$\Gammam_\ell = \text{diag}(\gamma^{(\ell)}_1,...,\gamma^{(\ell)}_{2^J})$.
Then, (\ref{matrix-channel}) can be written as
\beq
\Ym_\ell = \Am \Gammam_\ell^{1/2} \tilde{\Hm} + \Zm_\ell,
\label{eq:matrix_model2}
\eeq
where $\tilde{\Hm}\in \CC^{2^J\times M}$ with i.i.d. elements $\sim \Cc\Nc(0,1)$. 
Notice that in (\ref{eq:matrix_model2}) the number of total users $K_{\rm tot}$ plays no role. In fact, none of the matrices
involved in (\ref{eq:matrix_model2}) depends on $K_{\rm tot}$. 

\subsection{Inner decoding as a massive MIMO activity detection problem}

At this point, the inner decoder must simply identify which of the columns of $\Am$ is ``active'' in each subslot $\ell$. 
The set of detected active columns forms the list of submessages at subslot time $\ell$. 
The sequence of such lists is then passed to the outer decoder, which works as described in \cite{Ama2018} (see also 
Section \ref{sec:outercode}). Next, we discuss the detection of the active columns of $\Am$ in a given subslot $\ell$, 
and drop the subslot index for simplicity of notation since
such detection operation proceeds identically in each subslot and it is independent of the subslot index. 

Identifying the active columns of $\Am$ is tantamount estimating the diagonal elements of 
$\Gammam$, and identifying which ones are non-zero. This problem is completely analogous to the activity detection problem 
studied in \cite{liu2018massive,chen2018sparse,larsson2012piggybacking,senel2017device}. A difference here is that 
the elements of $\Gammam$ are random sums of the individual user channel gains $\{g_k\}$. 
Hence, even if the $g_k$'s were exactly individually known, or their statistics was known, 
these random sums would have unknown values and unknown statistics (unless averaging over all possible
active subsets, which would involve an exponential complexity in $K_{\rm tot}$ which is clearly infeasible in our context).
Hence, the Bayesian VAMP-based approaches advocated in  \cite{liu2018massive,chen2018sparse,larsson2012piggybacking,senel2017device}
do not find a straightforward application here. 
In contrast, we shall use our new non-Bayesian detection approach, that treats 
$\gammav := \diag(\Gammam) = (\gamma_1, \ldots, \gamma_{2^J})^\trasp$ as  a deterministically unknown vector. 
Interestingly, as already anticipated before, our algorithm is not only more robust and more generally applicable than Bayesian VAMP, 
but yields also better results for the estimation of $\gammav$ even under the assumption that 
the individual elements $\gamma_r$ are Bernoulli variables taking on the value zero with probability $(1 - 1/2^J)^{K_a}$ 
or a {\em known positive value} with probability $1 - (1 - 1/2^J)^{K_a}$. The reason is that, despite the nice features 
of Bayesian VAMP, such as the fact that their performance in the large dimensional limit, can be predicted in closed form 
by the {\em State Evolution} equation
\cite{bayati2011dynamics,kim2011belief,liu2018sparse} when 
the coding matrix of $\Am$ has sub-Gaussian i.i.d. elements, for the case where $n_0$ is small with respect to $M$, VAMP becomes
numerically unstable and sometimes produces completely wring estimates of the activity vector $\gammav$. 
Notice that in a practical application, the subslot dimension $n_0$ may be of the order of 100 to 400 symbols, while
for a city-wide IoT data collector it is not unreasonable to have $M$ of the order of 500 to 1000 antennas (especially when 
considering narrowband signals such as in LoRA-type applications
\cite{centenaro2016longrange,bankov2016limits}. 
This is precisely the regime where we have observed a critical behavior of VAMP, while our algorithm uniformly improves as
$M$ increases, for any slot dimension $n_0$. In Section \ref{MLalgo}, we give the details of the proposed 
inner decoding algorithm. 

\subsection{Outer Code} \label{sec:outercode}

For each subslot $\ell$, let $\widehat{\gammav}^{(\ell)} = (\widehat{\gamma}_1, \ldots, \widehat{\gamma}_{2^J})^\trasp$ 
denote the ML estimate of $\gammav_\ell$ obtained by the inner decoder. 
Then, the list of active messages at subslot $\ell$ is defined as
\begin{equation}
\label{support-detection} 
\Sc_\ell = \left \{ r \in [1:2^J] : \widehat{\gamma}_r^{(\ell)} \geq \tau_\ell \right \}, 
\end{equation}
where $\tau_1, \ldots, \tau_L$ are suitable pre-defined thresholds.  Let $\Sc_1, \Sc_2, \ldots, \Sc_L$ the sequence of lists of active subblock 
messages. Since the subblock (coded) messages contain parity bits with parity profile $\{0,p_2, \ldots, p_L\}$, 
not all all message sequences in $\Sc_1 \times \Sc_2 \times \cdots \times \Sc_L$ are possible. The role of the outer decoder is
to identify all possible subblock message sequences. 
The output list $\Lc$ is initialized as an empty list.
Starting from $\ell=1$ and proceeding in order, the decoder converts the integer indices 
$\Sc_\ell$ back to their binary representation, separates data and parity bits, computes the parity 
checks for all the combinations with messages from the list $\Lc$ and extends only the paths 
in the tree which  fulfill the parity checks. 
A precise analysis of the error probability of such decoder and its complexity in terms of 
surviving paths in the list is given in \cite{Ama2018}.

\section{Maximum Likelihood Activity Detection}  \label{MLalgo}

In \cite{Hag2018} we have addressed the estimation of $\gammav$ from the $n_0 \times M$ received signal matrix 
of the form \eqref{eq:matrix_model2}  using the Maximum Likelihood (ML) approach.
Given $\Am$ (known to the decoder), and $\Gammam$ (deterministic unknown), 
the columns $\yv_i$ of $\Ym = [\yv_1, \ldots, \yv_M]$ in \eqref{eq:matrix_model2} 
are i.i.d. Gaussian vectors $\yv_i \sim \mathcal{CN}(0, \Sigmam_\yv)$ with (conditional) covariance matrix
\begin{align}
    \Sigmam_\yv = \Am \Gammam \Am^\herm + N_0 \Id_{n_0}
    =\sum_{r=1}^{2^J} \gamma_r \av_r \av_r^\herm + N_0 \Id_{n_0}.
\label{eq:true_cov}
\end{align}
We also  define the empirical covariance matrix of the columns of $\Ym$ as
\begin{align}
    \widehat{\Sigmam}_\yv=\frac{1}{M} \Ym \Ym^\herm 
    =\frac{1}{M} \sum_{i=1}^M \yv_i \yv_i^\herm.
\label{eq:samp_cov}
\end{align}
The log-likelihood function is the logarithm of the 
conditional distribution of $\Ym$ given $\gammav$ (up to multiplicative and additive irrelevant factors). 
This takes on the form
\begin{align}
    f(\gammav):= -\frac{1}{M}\log p(\Ym| \gammav)
    &=-\frac{1}{M}\sum_{i=1}^M \log p(\yv_i| \gammav)\\
    &=  \log | \Am \Gammam \Am^\herm+ N_0 \Id_{n_0}| \nonumber \\
    &+ \text{tr}\left ( \Big( \Am \Gammam \Am^\herm+ N_0 \Id_{n_0} \Big ) ^{-1}
    \widehat{\Sigmam}_\yv \right).
    \label{eq:ML_cost}
\end{align} 
The ML estimate of $\gammav$ is given by 
\beq
\gammav^* = \argmin_{\gammav \in \mathbb{R}_+^{2^J}} f(\gammav).
\eeq
It is shown in \cite{Hag2018}, that $f(\gammav)$ has only global minima if
$K_a \gtrsim \mathcal{O}(n_0^2)$. In the same work, the low complexity coordinate
descent Algorithm \ref{alg:ml_coord} is derived.
Compared to \ref{alg:ml_coord}, we used the matrix-inversion lemma
\beq
(\Sigmam + d(\av \av^\herm))^{-1} = 
        \Sigmam ^{-1} +
        \frac{d\, \Sigmam^{-1} \av \av^\herm \Sigmam^{-1}}
        {1+ d \, \av^\herm \Sigmam^{-1} \av}
\eeq
to reformulate the algorithm in order to avoid explicit matrix inversion.
Note that the log-likelihood function \eqref{eq:ML_cost} depends on $\Ym$ 
only through the $n_0\times n_0$ empirical covariance matrix $\widehat{\Sigmam}_{\yv}$.
The calculation of the covariance matrix itself scales with $M$, but the minimization
of \eqref{eq:ML_cost} is independent of $M$. For large $M$, this gives a crucial
advantage in complexity compared to other algorithms like VAMP \cite{chen2018sparse}, whose complexity 
scales at least linear in $M$.
\begin{algorithm}
	\caption{Activity Detection via Coordinate Descend } 
	\label{alg:ml_coord} 
{
	\begin{algorithmic}[1]
		\State {\bf Input:}
        The sample covariance matrix $\widehat{\Sigmam}_\yv=\frac{1}{M} \Ym \Ym^\herm$
        of the $n_0 \times M$ received signal matrix $\Ym$.
		\State {\bf Initialize:}
        $\Sigmam^{-1}=\frac{1}{N_0} \Id_{n_0}$, $\gammav={\bf 0}$.
		\For { $i=1,2, \dots$}
		\State {Select an index $r \in [2^J]$ corresponding to the $r$-th component
        of  $\gammav=(\gamma_1, \dots, \gamma_{2^J})^\top$
        randomly or according to a specific schedule.}
		\vspace{2mm}
		\State {\bf ML:}
            Set 
            \beq
            d^*= \max \Big \{\frac{ \av_r^\herm \Sigmam^{-1}
            \widehat{\Sigmam}_\yv \Sigmam^{-1} \av_r -  \av_r^\herm \Sigmam^{-1}\av_r }
            {(\av_r^\herm \Sigmam^{-1}\av_r )^2}, -\gamma_{r} \Big \}
            \eeq
		\State Update $\gamma_{r} \leftarrow \gamma_{r}+ d^*$.
        \State Update
        \beq
        \Sigmam^{-1} \leftarrow \Sigmam ^{-1} +
        \frac{d^*\, \Sigmam^{-1} \av_r \av_r^\herm \Sigmam^{-1}}
        {1+ d^* \, \av_r^\herm \Sigmam^{-1} \av_r}
        \eeq
		\EndFor
		\State {\bf Output:}  The resulting estimate $\gammav$.
	\end{algorithmic}}
\end{algorithm}

\section{Discussion and Analysis}
\label{sec:analysis}

\begin{Figure}
    \centering
    \includegraphics[width=\linewidth]{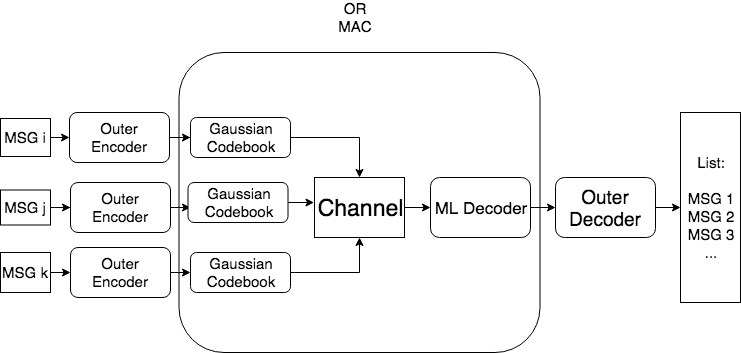}
    \captionof{figure}{ Proposed Coding Scheme }
    \label{fig:channel}
\end{Figure}

\subsection{Inner Code}
The results of \cite{Hag2018} suggest that the probability of an error in the estimation of
the support of $\gammav$ vanishes in the limit $M \to \infty$ for any SNR $\frac{P}{N_0} > 0$
as long as
$K_a = \mathcal{O}(n_0^2) = \mathcal{O}(\frac{n^2}{L^2})$. 
This is shown to hold also 
for $(M,K_a, \frac{P}{N_0}) \to (\infty, \infty, 0)$ as long as 
\beq
\frac{M}{\text{max}((\frac{N_0}{P})^2,K_a)} = o(1).
\label{eq:scaling}
\eeq
More precisely \cite{Hag2018} gives the following bound for the reconstruction
error of 
\begin{align}
\lVert \gammav - \gammav^*\rVert_2 \leq
\kappa \left ( \frac{(P/N_0)^{-1}}{\sqrt{M}} + \sqrt{\frac{K_a}{M}} \|\gammav\|_2 \right  )
\label{eq:gamma_perf}
\end{align}
where $\kappa$ is some universal constant and $\gammav^*$ denotes the estimate of
$\gammav$ by Non-Negative Least-Squares (NNLS) as proposed in \cite{wang2018fly}. We omit here the details of the 
NNLS scheme for the sake of brevity and since this is identical to what was proposed in \cite{wang2018fly} and analyzed in \cite{Hag2018}.
The numerical results in \cite{Hag2018} suggest that the reconstruction error of the ML algorithm
is at least as good as that of NNLS (in practice it is typically {\em much better}). 
Work in progress by the authors aims to shown that, under certain conditions, the reconstruction error bound (\ref{eq:gamma_perf}) 
applies also to ML. This bound is indeed very conservative and in addition it is difficult if not impossible to sharply quantify the constant $\kappa$. 
Nevertheless, this is enough to give achievable scaling laws for the probability of error of the inner decoder. 

\subsection{Outer Code}

In the case of error-free support recovery, the support $\sv^{(\ell)}$ of the estimated $\widehat{\gamma}^{(\ell)}$ can be interpreted as
the output of a vector ``OR'' multiple access channel (OR-MAC) where the inputs are the binary columns of the activity matrix $\Bm_\ell$, namely, 
$\bv^{(\ell)}_k : k \in \Kc_a$, and the output is given by 
\beq
\sv^{(\ell)} =  \bigvee_{k \in \Kc_a} \bv_k^{(\ell)}, 
\eeq
where $\bigvee$ denotes the component-wise binary OR operation. 
The output entropy of the vector OR-MAC of dimension $2^J$
is bounded by the entropy of $2^J$ scalar OR-MACs. The marginal distribution of the entries of $\sv^{(\ell)}$ 
is Bernoulli with $\PP(s^{(\ell)}_r = 0) = (1 - 1/2^J)^{K_a}$. Hence, we have
\begin{equation} \label{outputH}
H(\sv^{(\ell)}) \leq 2^J \Hc_2 ( (1 - 1/2^J)^{K_a}). 
\end{equation}
Since all users make use of the same code, letting $R_\text{out}$ denote the rate of the outer code,
we have that the number of information bits sent by the $K_a$ active users over a subslot is 
$b_\text{sum} = K_a J R_\text{out}$. Therefore, in order to hope for small probability of error a necessary condition is
\beq
    K_aJ R_\text{out} \leq 2^J\mathcal{H}_2((1-1/2^J)^{K_a}).
    \label{eq:sumrate}
\eeq
For large $2^J$, we can approximate
\beq
    \mathcal{H}_2((1-1/2^J)^{K_a}) \approx K_a (1 + J - \log K_a).
\eeq
Inserting this into \eqref{eq:sumrate} and solving for $K_a$ we get
\beq
    K_a \leq 2^{J(1-R_\text{out}) + 1}.
\eeq
Together with the limitation of the inner ML decoder we get:
\beq
K_a \leq \min \left(c\frac{n^2}{L^2},2^{J(1-R_\text{out}) + 1}\right)
\label{eq:au_bound}
\eeq
for some constant $c$. 

Based on this analysis, we can identify two operational regimes. 
For large enough block length, the number of active users
is limited by the outer code, while for small block lengths it is limited by the detection
capabilities of the inner decoder. 

Together with the considerations of the previous section we can estimate the order of the
number of antennas needed to achieve a certain value of $E_b/N_0$.
Let the number of active users $K_a$ be fixed and let the blocklength $n$ be large enough
such that the limiting factor in \eqref{eq:au_bound} is the outer code, i.e.,
\begin{equation} \label{regime-large-n}
c \frac{n^2}{L^2} \geq 2^{J(1-R_\text{out}) + 1}. 
\end{equation}
Then, we can write
\begin{align}
    \frac{E_b}{N_0} &= \frac{P}{N_0}\frac{n}{b} \\
                    &= \frac{P}{N_0}\frac{n}{LJR_\text{out}}\\
                    &\geq \frac{P}{N_0}
    \frac{2^{J(1-R_\text{out})/2 + 1/2}}{\sqrt{c}JR_\text{out}}
\end{align}
which is equivalent to
$\frac{P}{N_0} \leq \frac{E_b}{N_0} \phi(J, R_\text{out})$ with $\phi(J, R_\text{out}) =
\frac{\sqrt{c}JR_\text{out}}{2^{J(1-R_\text{out})/2 + 1/2}}$.
The bound \eqref{eq:gamma_perf} gives that $M$ should be of order
\beq
M = \Omega\left(\text{max}\left(\left(\frac{E_b}{N_0}\phi(J,R_\text{out})\right)^{-2}, K_a\right)\right).
\eeq
It is therefore apparent that for any $E_b/N_0 > 0$, $J$, and $R_\text{out}$, if (\ref{regime-large-n}) holds and 
$K_a \leq 2^{J(1-R_\text{out}) + 1}$, we can find a sufficiently large value of $M$ such that 
a desired level of reliability (in terms of the PUPE) is achieved. 

\section{Simulations}
\label{sec:numerics}

\subsection{Thresholding and Power Allocation}

The outer decoder  requires a hard decision on the support of the estimated $\widehat{\gammav}^{(\ell)}$. 
One approach consists of selecting the $K_a + \Delta$ largest entries in each section, where $\Delta \geq 0$
can be adjusted to balance between false alarm and misdetection in the outer channel. However, this approach requires the
the knowledge of $K_a$, which is generally unknown and a random variable itself. 
An alternative approach, which does not require this knowledge, consists of fixing a sequence of thresholds 
$\{\tau_\ell : \ell \in [1 : L]\}$ and let $\sv^{(\ell)}$ to be the binary vector of dimension $2^J$ 
with elements equal to 1 for all components $r \in \Sc_\ell$, where $\Sc_\ell$ is given in (\ref{support-detection}). 
By choosing the thresholds, we can balance between missed detections and false alarms. 
The outer decoder has the property that any missed detection will necessarily result in a missing message 
in the final output list $\Lc$, while a relatively large number of false alarms can be corrected.
The concrete number of correctable errors depends on the distribution of parity bits and
is mainly limited by the runtime of the decoder, since  many false alarms in the early subslots 
may let the runtime grow exponentially. It is found in \cite{Ama2018}, that the best performance of the outer tree decoder 
is achieved by choosing all of the parity bits in the last sections, but this increases the 
complexity of the outer decoder to an unfeasible level. Therefore,  a compromise has to be found between 
rate and complexity. We have found  that it is advantageous to use a non-uniform power allocation 
to reduce the complexity, i.e.,  instead of allocating the same power to the inner codewords of each subslot,
we choose a sequence $P_1, \ldots, P_L$ such that $\sum_{\ell=1}^L P_\ell = LP$. A decaying power allocation can reduce the number 
of false alarms in the early subslots at the cost of an increased number of false alarms more of them in the later sections. In this way, 
we can choose more parity bits in the later subslots, i.e., an increasing parity profile $p_2 \leq p_3 \leq \cdots \leq p_L$, 
while keeping the complexity manageable.

\subsection{Results}

For the simulations in \figref{fig:sim} we choose $b = 96$ bits as payload size for each user 
and $n = 3200$ complex channel uses. We choose $L = 32$, yielding $n_0 = 100$ and $J = 12$, yielding an inner coding rate
$R_{\rm in} = 12/100$ 
bits per channel use. 
Notice that the coding matrix $\Am$ in this case has dimension $100 \times 4096$ and therefore is still quite manageable. 
Notice also that if one wishes to send the same payload message using the piggybacking scheme of 
\cite{larsson2012piggybacking,senel2017device}, each user should make use of $2^{96}$ columns, which 
is totally impractical. 

For the outer code, we choose the following parity profile $p = [0,9,9, \ldots ,9,12,12,12]$, yielding an outer coding
rate $R_{out} = 0.25$ information bits per binary symbol. 
We fix $N_0 = 1$ and choose the transmit power, such that $E_b/N_0 = 0$dB. 
All large scale fading coefficients are fixed to $g_k \equiv 1$. \figref{fig:sim} shows the total PUPE 
$P_e = p_{md} + p_{fa}$ as a function of the number of active users for different numbers of receive antennas $M$.  
We find that with $M = 400$ the error probability drops below $10^{-2}$ for 300 active users.
Notice that this corresponds to a total spectral efficiency $\mu = \frac{12}{100} \times 0.25 \times 300 = 9$ bit per channel use, which is 
significantly larger than today's LTE cellular systems (in terms of bit/s/Hz per sector) and definitely much larger than IoT-driven schemes such as 
LoRA \cite{centenaro2016longrange,bankov2016limits}. 
According to the random coding bound of \cite{Polyanksiy:isit17} this is impossible to achieve for the scalar Gaussian channel 
(\ref{real-adder}) (only one receive antenna), even with coherent detection and roughly five times 
smaller spectral efficiency then here.

\begin{Figure}
    \centering
    \includegraphics[width=\linewidth]{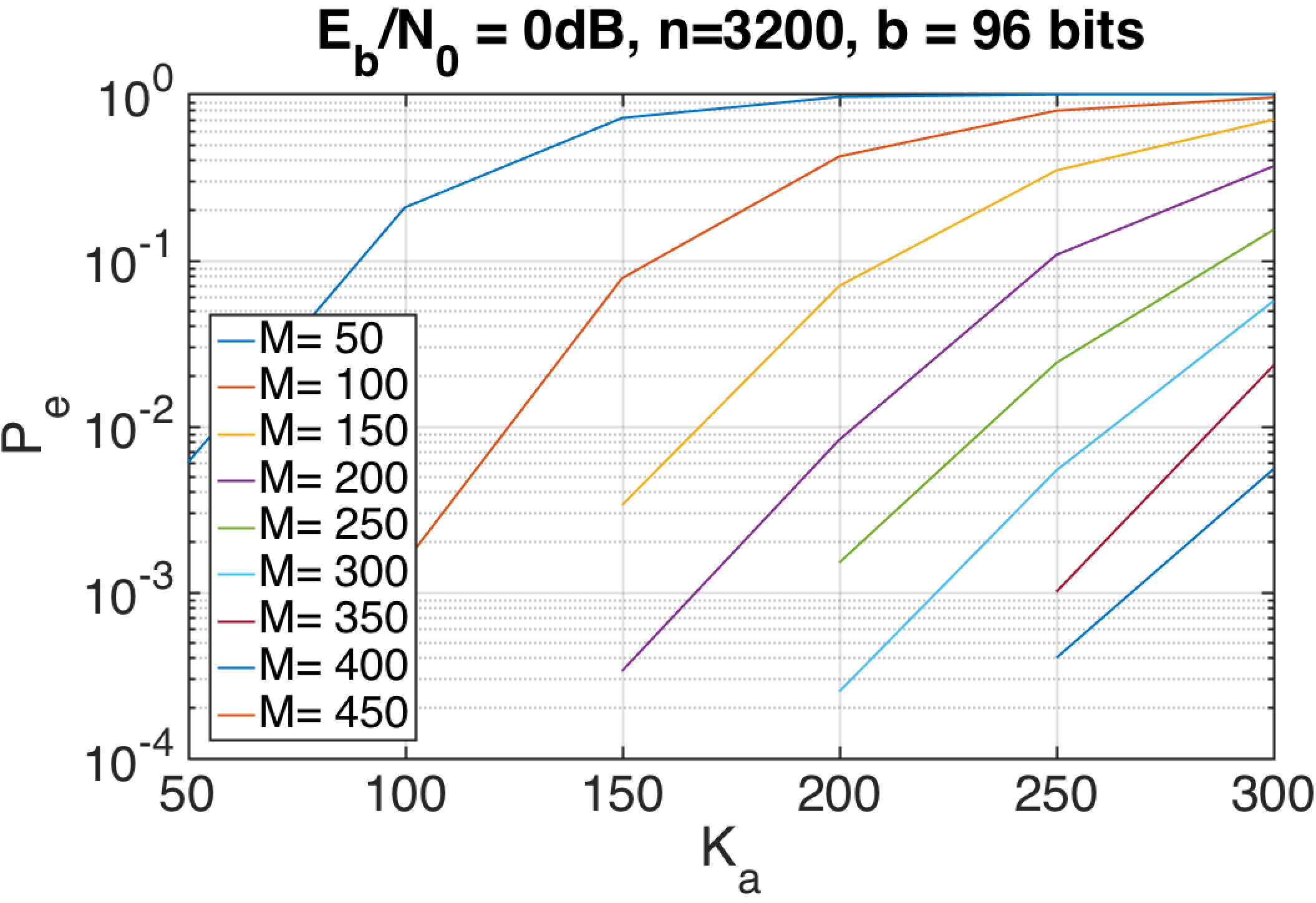}
    \captionof{figure}{ Error probability ($P_e = p_{md} + p_{fa}$) as a function of the number of
    active users for different numbers of receive antennas. $L = 32, J=12$. }
    \label{fig:sim}
\end{Figure}

\printbibliography
\end{document}

%% file: macros.tex
\long\def\comment#1{}

\newfont{\bbb}{msbm10 scaled 700}

\newfont{\bb}{msbm10 scaled 1100}
\newcommand{\CC}{\mbox{\bb C}}
\newcommand{\PP}{\mbox{\bb P}}
\newcommand{\RR}{\mbox{\bb R}}

\newcommand{\av}{{\bf a}}
\newcommand{\bv}{{\bf b}}

\newcommand{\ev}{{\bf e}}

\newcommand{\hv}{{\bf h}}

\newcommand{\sv}{{\bf s}}

\newcommand{\xv}{{\bf x}}
\newcommand{\yv}{{\bf y}}

\newcommand{\zerov}{{\bf 0}}

\newcommand{\Am}{{\bf A}}
\newcommand{\Bm}{{\bf B}}

\newcommand{\Gm}{{\bf G}}
\newcommand{\Hm}{{\bf H}}
\newcommand{\Id}{{\bf I}}

\newcommand{\Xm}{{\bf X}}
\newcommand{\Ym}{{\bf Y}}
\newcommand{\Zm}{{\bf Z}}

\newcommand{\Cc}{{\cal C}}

\newcommand{\Hc}{{\cal H}}

\newcommand{\Kc}{{\cal K}}
\newcommand{\Lc}{{\cal L}}

\newcommand{\Nc}{{\cal N}}

\newcommand{\Sc}{{\cal S}}

\newcommand{\gammav}{\hbox{\boldmath$\gamma$}}

\newcommand{\Gammam}{\hbox{\boldmath$\Gamma$}}

\newcommand{\Sigmam}{\hbox{\boldmath$\Sigma$}}

\newcommand{\diag}{{\hbox{diag}}}

\newcommand{\herm}{{\sf H}}
\newcommand{\trasp}{{\sf T}}